\documentclass{basi}
\usepackage{epsfig}
\usepackage{txfonts}
\begin{document}
\title[Recurrent activity in AGN]{Recurrent Activity in Active Galactic Nuclei}
\author[D.J. Saikia and M. Jamrozy]
       {D.J. Saikia$^{1,2}$\thanks{e-mail:djs@ncra.tifr.res.in, jamrozy@oa.uj.edu.pl} and M. Jamrozy$^3$ \\
      $^1$ National Centre for Radio Astrophysics, Post Bag 3, Ganeshkhind, Pune 411 007 \\
      $^2$ ICRAR, University of Western Australia, Crawley, WA 6009, Australia \\
      $^3$ Obserwatorium Astronomiczne, Uniwersytet Jagiello\'nski, ul. Orla 171, PL-30244 Krak\'ow, Poland \\
        }
\date{Received 30 December 2009; accepted 31 December 2009}
\maketitle
\label{firstpage}
\begin{abstract}
There has been a growing body of evidence to suggest that AGN activity, which is powered
by mass accretion on to a supermasive black hole, could be episodic, although the range
of time scales involved needs to be explored further. The structure and spectra of radio 
emission from radio galaxies, whose sizes range up to $\sim$5 Mpc, contain information
on the history of AGN activity in the source.  They thus provide a unique opportunity to study the 
time scales of recurrent AGN activity.  The most striking examples of recurrent activity in radio galaxies
and quasars are the double-double or triple-double radio sources which contain two or three
pairs of distinct lobes on opposite sides of the parent optical object. Spectral and dynamical
ages of these lobes could be used to constrain time scales of episodic activity. Inverse-Compton
scattered cosmic microwave background radiation could in principle probe   
lower Lorentz-factor particles than radio observations of synchrotron emission, and 
thereby reveal an older population. We review briefly the radio continuum as well as molecular and
atomic gas properties of radio sources which exhibit recurrent or episodic activity, and 
present a few cases of quasars which require further observations to confirm their episodic nature.
We also illustrate evidence of episodic AGN activity in radio sources in clusters of
galaxies.
\end{abstract}

\section{Introduction}  
\label{sec:intro}
It is widely believed that active galactic nuclei (AGN) are powered by mass accretion 
onto a supermassive black hole (SMBH) whose mass ranges from $\sim$10$^6$ to 
10$^{10}$ M$_\odot$ (e.g. Salpeter 1964; Zel'dovich \& Novikov 1964; Lynden-Bell 1969; Rees 1984).
The SMBHs have been inferred and their mass estimated via a number of techniques which include
stellar and gas dynamics in the circumnuclear regions of nearby galaxies, reverberation mapping 
methods for nearby and more distant AGN (e.g. Kormendy \& Richstone 1995; Kormendy \& Gebhardt 2001; 
Merritt \& Ferrarese 2001), and the kinematics of maser emitting regions seen at radio wavelengths  
(e.g. Miyoshi et al. 1995). Nearly every nearby galaxy appears to have an SMBH at its centre with 
the mass of the black hole correlated with some of the properties of the host galaxy such as
the spheroid or bulge luminosity and  mass (Kormendy \& Richstone 1995; Magorrian et al. 1998),
light concentration (Graham et al. 2001) and central stellar velocity dispersion (Ferrarese \&
Merritt 2000; Gebhardt et al. 2000). Using these relationships, Marconi et al. (2004) estimate
the local black hole mass function (BHMF), and find that the BHMF of AGN relics, which are black
holes grown entirely by mass accretion during AGN phases, to be consistent with the local BHMF.
They suggest a scenario in which the local black holes grew during AGN phases with the average
total lifetime of these active phases ranging from $\sim$1.5$\times$10$^8$ to 10$^9$ yr. 

Amongst the galaxies harbouring an AGN, a small fraction appear to be luminous at radio 
wavelengths, and there have been suggestions that the radio loudness may be episodic. 
In the SDSS (Sloan Digital Sky Survey) quasars, $\sim$8 per cent of the bright ones ($i<$18.5)
are radio loud in the sense that the ratio of radio to X-ray flux exceeds
unity (Ivezi\'{c} et al. 2002, 2004).
Nipoti, Blundell \& Binney (2005) studied the X-ray and radio emission from a few microquasars 
and noted the two distinct modes of energy output, namely the `coupled' mode in which the X-ray and 
radio luminosities are closely coupled and vary only weakly, and the `flaring' mode, in which the
radio luminosity increases dramatically with little impact on the X-ray luminosity. A typical 
microquasar is in the flaring mode, which involves relativistic jets, only a few per cent of the time. 
They noted that this is similar to the fraction of quasars that are radio loud, suggesting that the
essential difference between radio loud and radio quiet quasars may be the epoch at which a given
quasar is observed. 
K\"ording, Jester \& Fender (2006) also explored the possible similarities between AGN and X-ray binaries,
where the hardness-intensity diagrams have been used successfully to study the relationships between accretion
states and their connection to radio jets. 
Using a sample of 4963 SDSS quasars with ROSAT matches, they showed that an AGN is more likely to have a 
high radio to optical flux ratio when it has a high total luminosity or a large non-thermal contribution 
to the spectral energy distribution. Using also a sample of low-luminosity AGN and a simulated population
of X-ray binaries, they suggested that AGN and X-ray binaries have similar accretion states and associated
jet properties. Along with these ideas, Sikora, Stawarz \& Lasota (2007) suggested that producing 
radio-loud AGN via relativistic jets requires larger spin of the SMBH (see Blandford \& Zjanek 1977), and the
intermittency may be related to successful jet formation due to collimation by MHD outflows from the accretion
disks (see Begelman \& Li 1994). Another way to explain the difference involving stellar disruption by SMBHs 
has been explored (e.g. Gopal-Krishna, Mangalam \& Wiita 2008). However, models need to address the totality
of data on the radio loud vs. radio quiet dichotomy (e.g. Rafter, Crenshaw \& Wiita 2009).

\begin{figure}
\vbox{
     \psfig{file=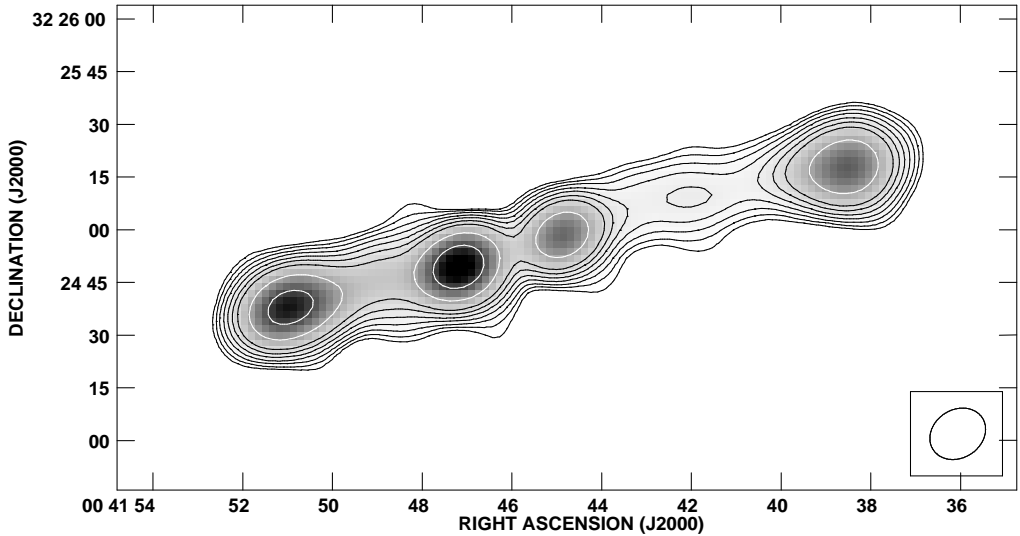,width=5.3in,angle=0}
      }
\caption[]{The GMRT image of the DDRG J0041+3224 at 240 MHz reproduced from Jamrozy et al. (2009).
The parent optical galaxy lies between the lobes of the inner double, 
which is slightly stronger than the outer double at 1400 MHz (Saikia, Konar \& Kulkarni 2006). Except 
for the sources with very compact doubles such as 3C236 and J1247+6723, the outer lobes tend to 
be more luminous than the inner ones.}
\end{figure}

\section{Radio-loud objects}
In the case of radio loud objects, the structure and spectral index distribution 
of the lobes of extended radio emission contain important information on the
history of the source, including radiative losses, reacceleration of particles
and possible episodic activity of the AGN. These have been used to study 
backflows from hotspots to form either the lobes and bridges of emission or 
possibly the X-shaped radio sources, tails of radio continuum emission due
to motion of the parent galaxy relative to the intracluster or intragroup 
medium, precession of jet axes as well as significant interruptions of jet
activity. Over the years several features in radio sources have been suggested
to be due to interruptions of jet activity. For example, a ridge of emisson
reminiscent of a jet but displaced towards the south of the nucleus in the 
radio galaxy 3C338 has been suggested to be possibly due to intermittent jet 
activity (Burns, Schwendeman \& White 1983). The radio galaxy 3C388 exhibits two 
distinct regions of emission separated by a jump in spectral index, which has been 
interpreted to be due to two different epochs of jet activity (Burns, Christiansen 
\& Hough 1982; Roettiger et al. 1994). They noted that this may be the first radio
galaxy in which multiple epochs of activity may be clearly visible in the large-scale
structure. Another suggestion of   a galaxy with 
distinct epochs of jet activity based on spectral index studies include
Her A, where the bright inner regions have flatter spectra with a reasonably sharp boundary
delineating it from the more extended lobe emission
(Gizani \& Leahy 2003; Gizani, Cohen \& Kassim 2005). The low-frequency observations
presented by Gizani, Cohen \& Kassim (2005) appear to confirm the multiple outburst interpretation
of the spectral index distributions.  In the radio galaxy 3C310 the inner
components B and D have substantially flatter spectra than the surrounding lobes
and could represent the most recent series of bubbles of plasma generated by the central
engine which have built up the radio source (van Breugel \& Fomalont 1984; Leahy, 
Pooley \& Riley 1986).  A lobe of emission
on only one side of the nuclear region has also been seen in the Gigahertz Peaked
Spectrum (GPS) radio source B0108+388, and has been suggested to be a relic of a previous
cycle of jet activity (Baum et al. 1990). However, the one-sidedness of the emission
is puzzling (cf. Stanghellini et al. 2005), and it would be interesting to examine
whether a `bursting bubble' model suggested for the northern middle lobe or NML 
(Morganti et al. 1999) may also be applicable in this source. Using the technique of
interplanetary scintillation observations with the Ooty Radio Telescope at 327 MHz,
Jeyakumar et al. (2000) identified a few possible candidates with low-frequency emission
on scales larger than the known milliarcsec-scale structure, but these need to be
studied further using Very Long Baseline Interferometry (VLBI) techniques. In a model 
to explain the abundance and size distribution of small radio sources, Reynolds \& Begelman (1997)
suggest that radio sources may be intermittent on time scales of $\sim$10$^4$$-$10$^5$ yr. 
There has also been
a suggestion of intermittent jet activity on different time scales in the quasar 3C273,
which has an apparently one-sided radio structure (Stawarz 2004).

\begin{figure}
      \hbox{
            \psfig{file=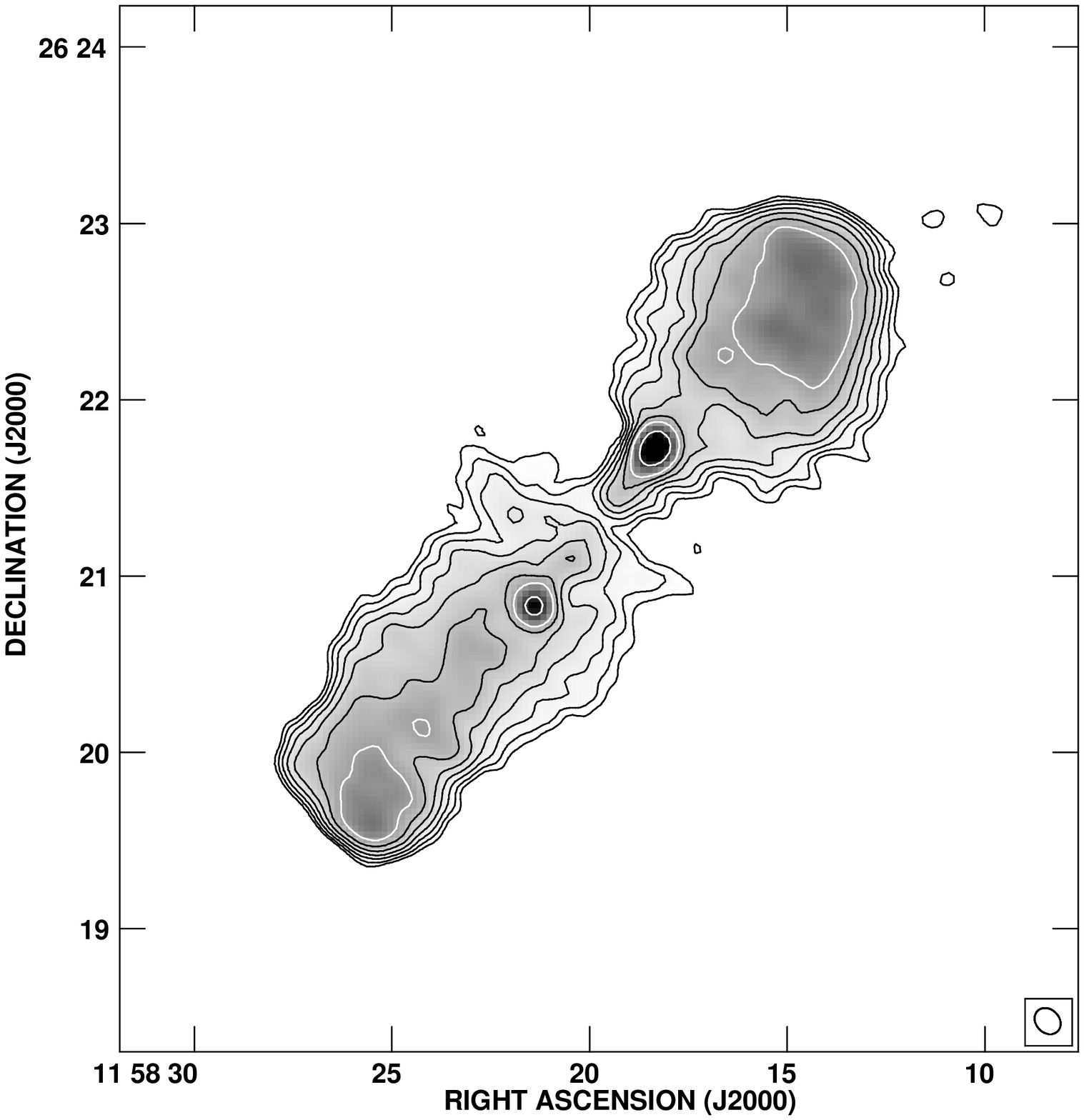,width=2.9in,angle=0}
            \psfig{file=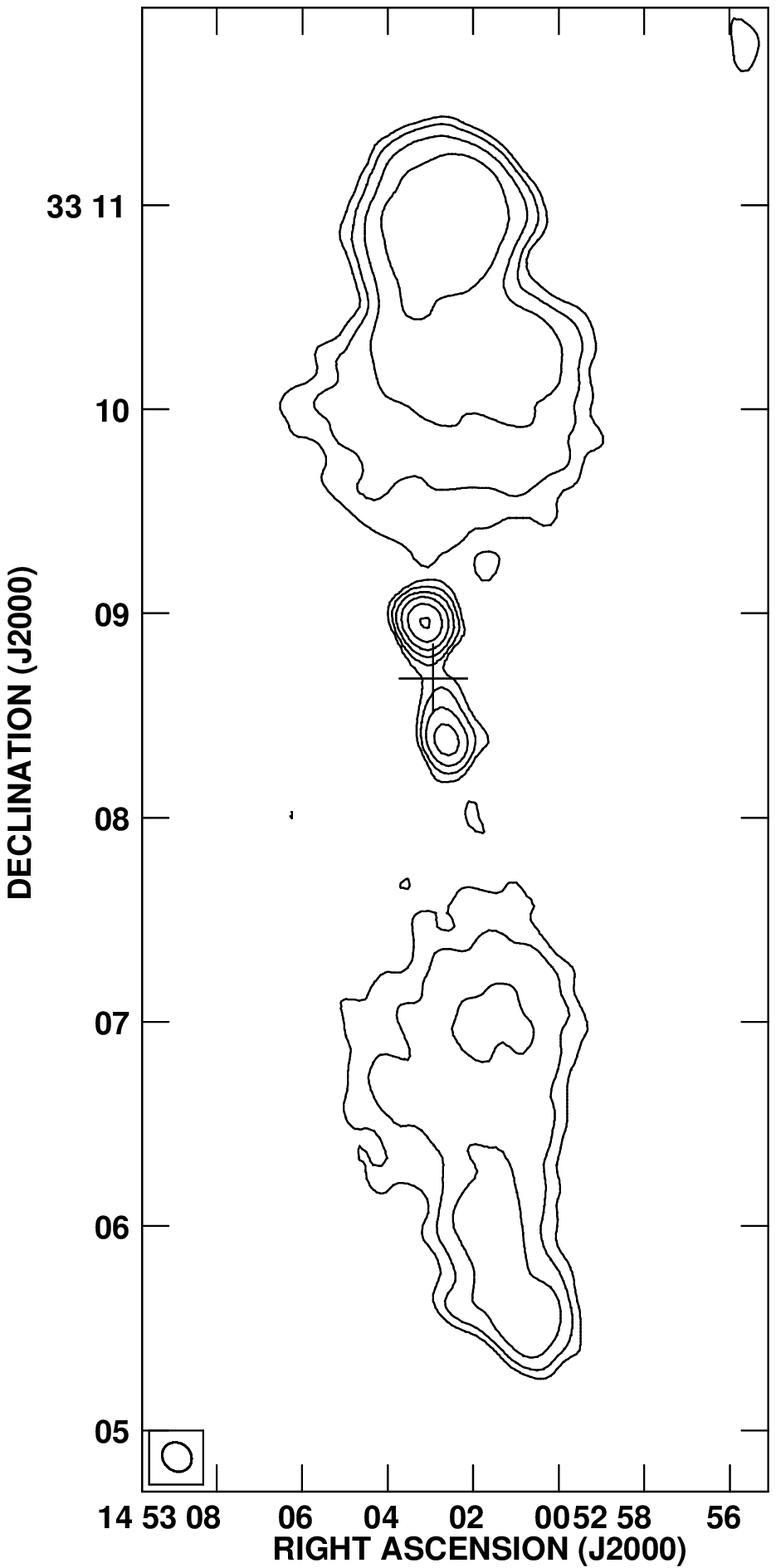,width=2.4in,angle=0}
            }
\caption[]{The GMRT images of the DDRGs J1158$+$2621 (left panel)
and J1453+3308 (right panel) at $\sim$335 MHz, reproduced from Jamrozy et al. (2009) and Konar et al. (2006) respectively.}
\end{figure}

A very striking example of episodic jet activity is when a new pair of radio lobes
is seen closer to the nucleus before the `old' and more distant radio
lobes have faded (e.g. Subrahmanyan, Saripalli \& Hunstead 1996; Lara et al. 1999). 
Such sources have been christened as 
`double-double' radio galaxies (DDRGs) by Schoenmakers et al. (2000a).
They proposed a relatively general definition of a DDRG as a double-double radio
galaxy consisting of a pair of double radio sources with a common centre, and also
suggested that the two lobes of the inner double should have an edge-brightened
radio morphology to distinguish it from knots in a jet. In such sources the newly-formed
jets propagate outwards through the cocoon formed by the earlier cycle of activity rather
than the general intergalactic or intracluster medium, after traversing through the
interstellar medium of the host galaxy. Approximately a dozen or so good examples of
such DDRGs are known.  These include objects such as 3C236
(Schilizzi et al. 2001) and J1247+6723 (Marecki et al. 2003) where the sizes of the inner
doubles are 1.7 kpc and 14 pc respectively to those such as J1835+6204 (Lara et al. 1999;
Schoenmakers et al. 2000b), where the size of the inner double could range up to several
hundred kpc. An interesting and well-studied example of a DDRG is Cen A, where in
addition to the diffuse outer lobes there are the more compact inner lobes and
a northern middle lobe (Burns, Feigelson \& Schreier 1983; Clarke, Burns \& Norman
1992; Junkes et al. 1993). Morganti et al. (1999) have detected a large-scale jet
connecting the northern lobe of the inner double and the NML, and have suggested that
the formation of the NML may be due to plasma which accumulated in the inner lobe 
and bursts out through a nozzle. Saxton, Sutherland \& Bicknell (2001) model the NML as a
buoyant bubble of plasma deposited by an intermittently active jet, while
Gopal-Krishna \& Wiita (2010) attribute the NML 
to interaction of the northern jet with a gaseous cloud associated with a stellar shell.
A recent image of Cen A made by Ilana Feain and her collaborators
contains suggestions of multiple cycles of activity (Feain, private communication). 

\begin{figure}
      \hbox{
            \psfig{file=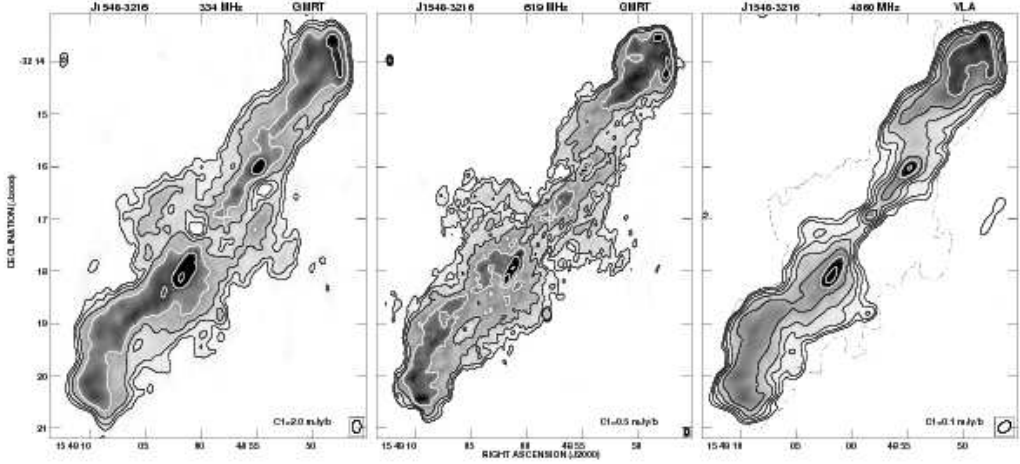,width=5.4in,angle=0}
            }
\caption[]{The images of the DDRG J1548$-$3216 at 334 and 619 MHz with the GMRT and at 4860 MHz
           with the Very Large Array (VLA) reproduced from Machalski, Jamrozy \& Konar (2010).  The cross marks the
           position of the optical galaxy.
           }
\end{figure}

\begin{figure}
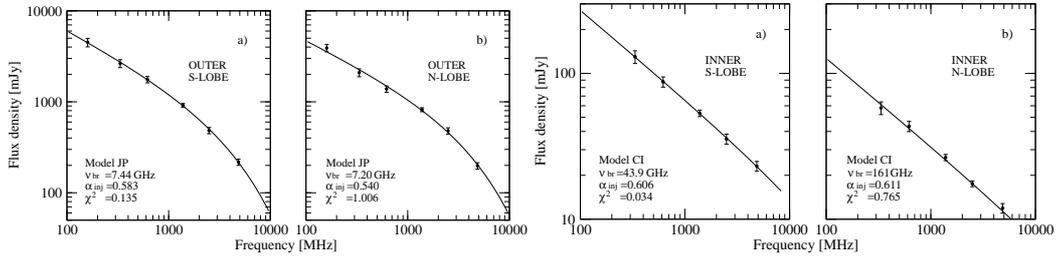

      \vbox{
            \psfig{file=12797f6.ps,width=2.7in,angle=0}
            \psfig{file=12797f9.ps,width=2.7in,angle=0}
            }
\caption[]{The spectra of the outer (left panels) and inner (right panels) lobes of the 
            DDRG J1548$-$3216 fitted with the Jaffe \& Perola (1973) and the continuous
            injection (Kardashev 1962; Pacholczyk 1970) models. These fits are
            from Machalski, Jamrozy \& Konar (2010) and have been made using the {\tt SYNAGE}
            package (Murgia 1996). While the outer lobes show evidence of
            ageing due to radiative losses, the spectra of
            the younger inner lobes are straight over the observed frequency range. 
          }
\end{figure}

\begin{table}[h!]
\caption{Sources with evidence of recurrent activity}
\begin{tabular}{ll  c  l  r r  l l }
\hline
Source       & Alt.        & Opt.    & Red-     & $l_{in}$   & $l_{o}$&    Notes               &   References \\
             & name        & Id.     & shift    &  kpc       &    kpc  &                       &              \\
%
\hline
J0041+3224   &  B2 0039+32  &  G  &   (0.45)     & 171        & 969    & DDRG                  &     1   \\
J0116$-$4722 &  PKS 0114$-$47 &  G  &   0.1461     & 460        &1447    & DDRG                  &     2   \\
J0821+2117   &  TXS 0818+214&  G  &   (1.0)      & 5.4        & 547    & DDRG                  &     19,20 \\
J0840+2949   &  4C29.30     &  G  &   0.0647     &  36        & 639    & DDRG                  &     18  \\
J0921+4538   &  3C219       &  G  &   0.1744     &  69        & 433    & DDRG                  &     3,4,5\\
J0929+4146   &              &  G  &   0.3650  &$\sim$30, 652  &1875    & TDRG                  &     6,39   \\
J0935+0204   &  4C02.27     &  Q  &   0.6491     &  70        & 470    & DDRQ                  &     16  \\
J1006+3454   &  3C236       &  G  &   0.1005     & 1.7        &4249    & DDRG                  &     7,8,9\\
J1158+2621   &  4C26.35     &  G  &   0.1121     & 138        & 483    & DDRG                  &     10  \\
J1242+3838   &              &  G  &   0.3000     & 251        & 602    & DDRG                  &     6   \\
J1247+6723   &  VII Zw 485  &  G  &   0.1073     &0.014       &1195    & DDRG                  &     11,12  \\
J1325$-$4301 &  Cen A       &  G  &   0.0018     &$\sim$12  &$\sim$600 & DDRG                  &     28,29,30,31,32,40 \\
J1352+3126   &  3C293       &  G  &   0.0450     & 1.6        & 190    & misaligned DDRG       &     36,37,38 \\
J1406+3411   &  3C294       &  G  &   1.7790     &            & 126    & relic X-ray           &     45,46 \\
J1453+3308   &  4C33.33     &  G  &   0.2481     & 159        &1297    & DDRG                  &     6,17   \\
J1504+2600   &  3C310       &  G  &   0.0538     &$\sim$90    & 320    & Flatter-$\alpha$ bubbles &  25,26  \\
J1548$-$3216 &  PKS 1545$-$321&  G  &   0.1082     & 313        & 961    & DDRG                  &     13,35,41,43  \\
J1651+0459   &  Her A       &  G  &   0.1540     &            & 513    & Steep-$\alpha$ relic  &     23,24 \\
J1835+6204   &  8C 1834+620 &  G  &   0.5194     & 369        &1379    & DDRG                  &     6,27,44   \\
J1844+4533   &  3C388       &  G  &   0.0917     &          &$\sim$70  & Steep-$\alpha$ relic  &     21,22 \\
J1959+4044   &  Cyg A       &  G  &   0.0561     &            & 136    & relic X-ray jet       &     33,34,42 \\
J2223$-$0206 &  3C445       &  G  &   0.0562     & 130        & 612    & DDRG                  &     14,15 \\
\hline
\hline
\end{tabular}
1: Saikia, Konar \& Kulkarni (2006); 2: Saripalli, Subrahmanyan \& Udaya Shankar 2002; 3: Perley et al. 1980;
4: Bridle, Perley \& Henriksen 1986; 5: Clarke et al. 1992;  6: Schoenmakers et al. 2000a;
7: Willis, Strom \& Wilson 1974;   8: Strom \& Willis 1980;   9: Schilizzi et al. 2001; 10: Owen \& Ledlow 1997;
11: Marecki et al. 2003;   12: Bondi et al. 2004;  13: Saripalli, Subrahmanyan \& Udaya Shankar 2003;
14: Kronberg, Wielebinski \& Graham 1986; 15: Leahy et al. 1997; 16: Jamrozy, Saikia \& Konar (2009);
17: Konar et al. (2006); 18: Jamrozy et al. (2007); 19: Marecki \& Szablewski (2009); 
20: Marecki et al. (2006); 21: Burns, Christiansen \& Hough (1982); 22: Roettiger et al. (1994);
23: Gizani \& Leahy (2003); 24: Gizani, Cohen \& Kassim (2005); 25: van Breugel \& Fomalont (1984);
26: Leahy, Pooley \& Riley (1986); 27: Lara et al. (1999); 28: Burns, Feigelson \& Schreier (1983); 
29: Clarke, Burns \& Norman (1992); 30: Junkes et al. (1993); 31: Morganti et al. (1999); 32: Feain et al.
(private communication); 33: Steenbrugge, Blundell \& Duffy (2008); 34: Streenbrugge, Heywood \& Blundell (2010);
35: Machalski, Jamrozy \& Konar (2010); 36: Akujor et al. (1996); 37: Bridle, Fomalont \& Cornwell (1981); 
38: Beswick et al. (2004); 39: Brocksopp et al. (2007); 40: Combi \& Romero (1997); 41: Subrahmanyan, Saripalli \&
Hunstead (1996); 42: Dreher, Carilli \& Perley (1987); 43: Safouris et al. (2008); 44: Schoenmakers et al. (2000b);
45: Erlund et al. (2006); 46: Blundell (2008).
\end{table}

Another interesting way of probing relic emission is via both X-ray and radio observations.
X-ray observations of inverse-Compton-scattered cosmic microwave background (ICCMB) radiation 
probe lower Lorentz-factor particles than radio observations of synchrotron emission, and 
therefore could in principle reveal an older population. Co-adding archival Chandra ACIS-I
data of Cygnus A, Steenbrugge, Blundell \& Duffy (2008) find evidence of a relic counter-jet indicating
a previous cycle of jet activity prior to the current jet activity which is seen as 
synchrotron radio emission. The non-detection of X-ray emission associated with any approaching
relic jet has been used to constrain the time scale between successive episodes of jet 
activity to $\sim$10$^6$ yr. More recently, Steenbrugge, Heywood \& Blundell (2010) have 
reported evidence of a relic counter lobe possibly energized by compression from the current
lobe, and hints of a relic hotspot at the end of the relic X-ray jet on the receding side.
Another interesting case is that of the powerful radio galaxy 3C294.  
Erlund et al. (2006) found that there is considerable extended X-ray emission, with the bulk of 
the X-ray emission being extended along an axis differently oriented from the radio axis by $\sim$50$^\circ$, 
and slightly longer than the length of the radio axis. Erlund et al. suggest that the offset between these 
two axes arises because of the precession of the axis along which jet plasma is ejected. 
The two discrete directions seen from a superposition of the radio and X-ray observations
reveal intermittent jet activity, one of them being traced by the freshly accelerated high-energy 
radio-synchrotron emitting particles while the other is consistent with ICCMB photons (see also Blundell 2008
for a review).

It is important to identify more examples of such sources for understanding episodic jet activity 
and constraining their time scales, and also for studying the propagation of jets in different media. 
For example, to explain the edge-brightened hotspots in the inner doubles of the DDRGs, Kaiser, 
Schoenmakers \& R\"{o}ttgering (2000) have suggested that warm (T$\sim$10$^4$K) clouds of gas in 
the intergalactic medium are dispersed over the cocoon volume by surface instabilities induced 
by the passage of the cocoon material. Although such features are not commonly seen, a systematic
and sensitive low-frequency survey of radio sources larger than a few hundred kpc or so is required
to clarify the frequency of occurrence of relic lobes or emission from earlier cycles of activity. 
A number of searches for such features, usually at $\sim$1400 MHz or higher have not yielded clear 
and striking examples of relic emission around bright radio sources (e.g. Reich, Stute \& Wielebinski 1980; 
Stute, Reich \& Kalberla 1980; Perley, Fomalont \& Johnston 1982; Kronberg \& Reich 1983; van der Laan, 
Zieba \& Noordam 1984; Jones \& Preston 2001). A deep multi-frequency survey at 153, 244, 610 and
1260 MHz towards a field containing clusters of galaxies was used to examine the structure and spectra
of 374 sources for evidence of recurrent or episodic AGN activity (Sirothia et al. 2009a). No
unambiguous case was found, suggesting that such activity is not commonly seen even at low radio
frequencies. However, most of the sources observed by Sirothia et al. (2009a) are small, with a median
angular size less than about 10 arcsec, which corresponds to a linear size of only $\sim$60 and 80
kpc at redshifts of 0.5 and 1 respectively.

\section{Notes on the sources}
The radio sources with evidence of recurrent activity are listed in Table 1, along with their optical
identification, redshift, projected linear size of the inner and outer doubles in kpc for those classified
as DDRGs, double-double radio quasar (DDRQ) 4C02.27, and the triple-double radio galaxy (TDRG), J0929+4146. 
In the latter case, the sizes of both the middle and inner doubles have
been listed. The sizes have been estimated using a cosmology with  
H$_0$=71 km s$^{-1}$ Mpc$^{-1}$, $\Omega_{\rm m}$=0.27, $\Omega_\Lambda$=0.73 (Spergel et al. 2003)
and the web-based cosmology calculator (Wright 2006).
 The list includes sources where there is reasonable evidence
for more than one cycle of activity, even if the source does not have two or more pairs of distinct
radio lobes. The references are not exhaustive but list those which are relevant for clarifying its
recurrent AGN activity. 
 
\noindent
{\bf J0041+3224:} A DDRG where the inner lobes are marginally brighter than the outer ones 
at 1400 MHz. The spectral indices of the components of the outer double are steeper than those
of the inner double, and the former exhibits some evidence of spectral steepening above $\sim$5000 MHz
which needs to be confirmed from lower-resolution, high-frequency observations  (Saikia, Konar \& Kulkarni 2006).   
The redshift also needs to be spectroscopically determined. For an estimated redshift of 0.45,
the size of the inner double is 171 kpc, well beyond the extent of the optical galaxy. A low-frequency
Giant Metrewave Radio Telescope (GMRT) image of the source is shown in Fig. 1.

\noindent
{\bf J0016$-$4722:} In addition to the two pairs of lobes, there is magaparsec-long radio feature
that is oriented perpendicular to the radio axis and has a high fractional polarization
(Saripalli, Subrahmanyan \& Udaya Shankar 2002).

\noindent
{\bf J0821+2117:} European VLBI Network (EVN) along with Multi-Element Radio-Linked Interferometer 
Network (MERLIN) observations of the central source at 1658 MHz show it to be a double with an angular
separation of 0.67 arcsec (Marecki \& Szablewski 2009), in addition to the outer lobes
which they estimate to have a separation of 68 arcsec (Marecki et al. 2006). Marecki et al. (2006)
estimate the redshift to be $\sim$1.0 based on the $r-z$ Hubble relation. As in the case of
3C236 and J1247+6723, where the inner doubles are less than a few kpc in size, the flux
density of the inner double is larger than the outer one. In this case, nearly 80 per
cent of the total flux density at 1400 MHz is from the inner double (see Marecki et al. 2006).

\noindent
{\bf J0840+2949:} The extended large-scale structure is reminiscent of the lobes in the earlier
cycle of activity, and hence has been called a DDRG. The spectral age of the inner double has
been estimated to be less than 33 Myr, while that of the extended diffuse emission is larger
than $\sim$200 Myr (Jamrozy et al. 2007).
 
\noindent
{\bf J0921+4538:} The edge-brightened strucuture of the north-western component of the inner
double and sharp boundary to the jet towards the south-east (e.g. Bridle, Perley \& Henriksen
1986; Clarke, Burns \& Norman 1992) has led us to presently classify it as a DDRG. 

\noindent
{\bf J0929+4146:} This source was reported to be a DDRG by Schoenmakers et al. (2000a) and
was later shown to have three distinct episodes of AGN activity and has been christened as
a triple-double radio galaxy by Brocksopp et al. (2007).
Kaiser, Schoenmakers \& R\"ottgering (2000) estimate the age of the 
middle and outer doubles to be 16 and 281 Myr respectively.

\noindent
{\bf J0935+0204:}
This is the only quasar so far with suggestions of a double-double structure (Jamrozy,
Saikia \& Konar 2009). The lobes of both the inner and outer doubles are highly asymmetric,
especially in flux density, and it is possible that the jets may also be intrinsically
asymmetric. The existence of a hot-spot in one of the lobes of the outer double indicates
that the time scale of episodic activity is less than a few Myr.

\noindent
{\bf J1006+3454:} This is one of the largest giant radio sources, and has a double-lobed
compact steep-spectrum source in the central region with a size of 1.7 kpc. The lobes of the
inner double are significantly more asymmetric than the outer ones in both location and 
flux density. The circumnuclear region shows evidence of star formation and 
H{\sc i} absorption against a lobe of the inner radio source (Conway \& Schilizzi 2000; 
Schilizzi et al. 2001; O'Dea et al. 2001). 

\noindent
{\bf J1158+2621:} A low-frequency GMRT image of this source, which was identified as a
DDRG from the survey of Abell clusters of galaxies by Owen \& Ledlow (1997) is shown in Fig. 2.
Kaiser, Schoenmakers \& R\"ottgering (2000) estimate the age of the 
inner and outer doubles to be 3.9 and 125 Myr respectively.

\noindent
{\bf J1242+3838:} Kaiser, Schoenmakers \& R\"ottgering (2000) estimate the age of the 
inner and outer doubles to be 12.8 and 184 Myr respectively.

\noindent
{\bf J1247+6723:} The central component is a GPS source (see Saikia, Gupta \& Konar 2007)
which is resolved into a compact double with a separation of 14 pc (Marecki et al. 2003;
Bondi et al. 2004). Saikia, Gupta \& Konar (2007) reported the detection of H{\sc i} in absorption
towards the GPS object. The absorption profile has multiple components and VLBI-scale
spectroscopic observations are required to identify the location of the different 
absorption components.

\noindent
{\bf J1325$-$4301:} This nearby radio galaxy, located $\sim$3.7 Mpc away (Hui et al. 1993; Ferrarese et al. 2007), 
exhibits two inner lobes, a northern middle lobe, and diffuse outer lobes subtending an angle of $\sim$10$^\circ$
in the sky (e.g.  Clarke, Burns \& Norman 1992; Junkes et al. 1993; Morganti et al. 1999). A more
recent image shows shell-like structures in the outer lobes (Feain et al., in preparation) which may be due to
earlier cycles of activity, in addition to the two well-defined inner and outer doubles. The molecular 
gas in the circumnuclear region (r$<$200 pc) as traced
by the CO(2$-$1) line is in the form of a disk or torus (Espada et al. 2009), perpendicular to 
the X-ray/radio jet (Kraft et al. 2003, 2007). Espada et al. (2009) suggest a close connection between
the molecular gas and the AGN. Broad H{\sc i}  absorption has been detected towards the central  
region (e.g. Sarma, Troland \& Rupen 2002; Morganti et al. 2008), and  
ascribed to a cold, circumnuclear disk (Morganti et al. 2008). The overall spectra of the 
large outer lobes steepen at frequencies above 5 GHz, but with significant differences for the 
northern and southern lobes, possibly due to the influence of the northern middle lobe
(Israel et al. 2008; Hardcastle et al. 2009). Between 80 MHz and 43 GHz, the spectra of the 
north-eastern and south-western inner lobes are fitted well by a power law of spectral index 
(S$\propto\nu^{-\alpha}$), $\alpha$$\sim$0.7
(Alvarez et al. 2000, and references therein). Further details on Cen A can be found in the
proceedings of the conference `Many faces of Centaurus A' which is scheduled to appear in
Publications of the Astronomical Society of Australia.

\noindent
{\bf J1352+3126:} The size of $\sim$1.6 kpc for the inner double is between the two prominent
peaks of emission, which are edge brightened, and appears to have a cirumferential field at
the outer edges (Akujor et al. 1996), characteristic of hotspots (see Saikia \& Salter 1988
for  a review). There are diffuse blobs of emission beyond both the hotspots of the inner
double. The inner double contributes more than 70 per cent of the total flux density of the
source and is misaligned by $\sim$35$^\circ$ from the axis of the large-scale structure. 
Evans et al. (1999) reported it to be rich in molecular gas, and from multiwavelength imaging suggested
it to be a disk galaxy with an optical jet/tail extending towards an apparent companion galaxy. 
They suggested that the radio activity may have been triggered by a gas-rich galaxy-galaxy interaction 
or merger event. Neutral hydrogen absorption features which are both blue- and red-shifted relative
to the systemic velocity (Beswick, Pedlar \& Holloway 2002; Beswick et al. 2004), and fast outflowing gas
blue-shifted by upto $\sim$1000 km s$^{-1}$ (Morganti et al. 2003; Emonts et al. 2005) have been reported.

\noindent
{\bf J1406+3411:} The radio and X-ray images, tracing relatively freshly accelerated radio-synchrotron
emitting particles and ICCMB photons respectively, are misaligned by $\sim$50$^\circ$, suggesting 
intermittent AGN activity in a source with a precessing jet (Erlund et al. 2006; Blundell 2008).

\noindent
{\bf J1453+3308:} The maximum spectral ages for the northern and southern lobes of the outer double have
been estimated to be $\sim$47 and 58 Myr, respectively, while the spectrum of the inner double is practically
straight with an age less than about 2 Myr (Konar et al. 2006). Kaiser, Schoenmakers \& R\"ottgering (2000) 
estimate the dynamical ages of the inner and outer doubles to be 1.6 and 215 Myr respectively. A low-frequency
GMRT image of this source is shown in Fig. 2.

\noindent
{\bf J1504+2600:} The integrated spectral index of the source was estimated by Laing \& Peacock (1980)
to be $\sim$1.3 which is unusually steep for sources of similar luminosity, suggesting that it is due to
old or relic emission. The lobes show considerable substructure, of which the components B and D on 
opposite sides of the nucleus appear to have significantly flatter spectra than the surroundings, 
suggesting that they represent the most recent AGN activity (Leahy, Pooley \& Riley 1986).

\noindent
{\bf J1548$-$3216:} Safouris et al. (2008) present detailed high-resolution images of this DDRG,
which was reported by Saripalli, Subrahmanyan \& Udaya Shankar (2003),  and suggest that 
the interruption to jet activity has been less than a few per cent of the lifetime of 
$\sim$(0.3$-2)\times$10$^8$ yr of the giant radio source.  From a multi-frequency study of this DDRG, 
Machalski, Jamrozy \& Konar (2010) estimate the dynamical ages of the outer and the inner doubles to 
be 132$\pm$28 Myr and 9$\pm$4 Myr, respectively. They find the  radiative age of the oldest plasma in the outer 
lobes is $\sim$65$-$75 Myr towards the centre of the old cocoon, and $\sim$5$-$15 Myr for the inner double.
The GMRT low-frequency images and spectra of the outer and inner doubles from Machalski, Jamrozy \& Konar (2010)
are shown in Figs. 3 and 4 respectively.

\noindent
{\bf J1651+0459:} The possibility of two cycles of activity has been inferred from a somewhat sharp boundary
in the spectral index distribution with the bright inner regions having flatter spectra 
from the more extended lobe emission (Gizani \& Leahy 2003; Gizani, Cohen \& Kassim 2005).

\noindent
{\bf J1835+6204:}
Kaiser, Schoenmakers \& R\"ottgering (2000) estimate the dynamical ages of the inner and outer doubles to
be 5.4 and 117 Myr respectively. Schoenmakers et al. (2000b) present a detailed study of this source and
use the presence of a hotspot in the northern outer lobe, to constrain the advance velocity of the inner
lobes to be in the range 0.19$-$0.29$c$, which corresponds to an age of 2.6$-$5.8 Myr.

\noindent
{\bf J1844+4533:}
The radio galaxy appears to exhibit two distinct regions of emission delineated by a 
jump in the spectral index, which has been suggested to be due to two different cycles of AGN
activity (Burns, Christiansen \& Hough 1982; Roettiger et al. 1994). 

\noindent
{\bf J1959+4044:} Steenbrugge, Blundell \& Duffy (2008) have reported evidence of a relic
counter-jet from X-ray observations, and have estimated a time scale of $\sim$10$^6$ yr 
between successive episodes of jet activity. More recently, 
possible evidence of a relic counter lobe has been suggested by Steenbrugge, Heywood \& Blundell (2010). 

\noindent
{\bf J2223$-$0206:} The structure of the inner double is reminiscent of 3C219, with a jet-like, edge-brightened
 structure towards the south and an edge-brightened hotspot towards the north (Leahy et al. 1997). A lower-resolution
image has been published by Kronberg, Wielebinski \& Graham (1986). 

\begin{figure}
      \hbox{
            \psfig{file=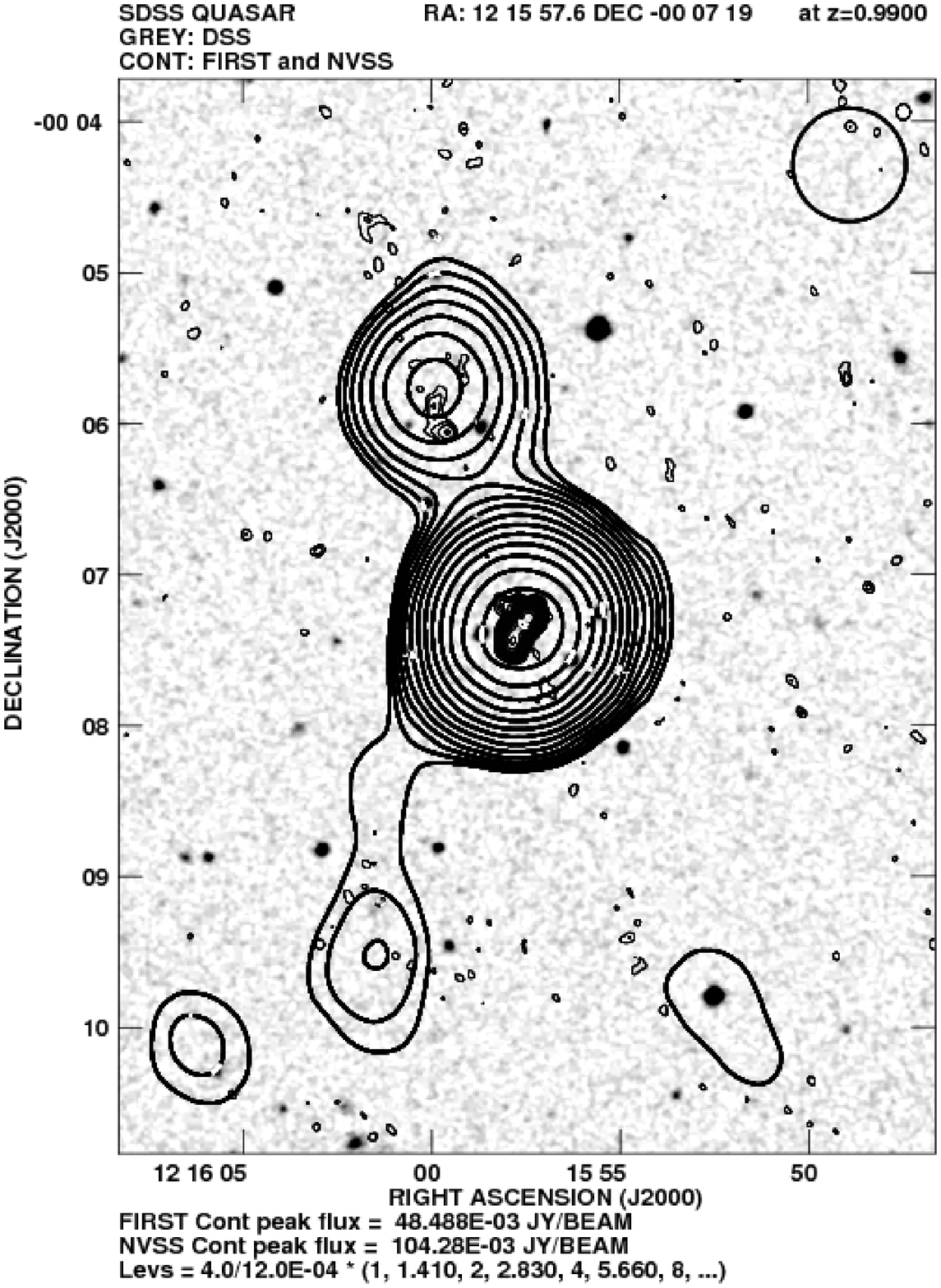,width=2.7in,angle=0}
            \psfig{file=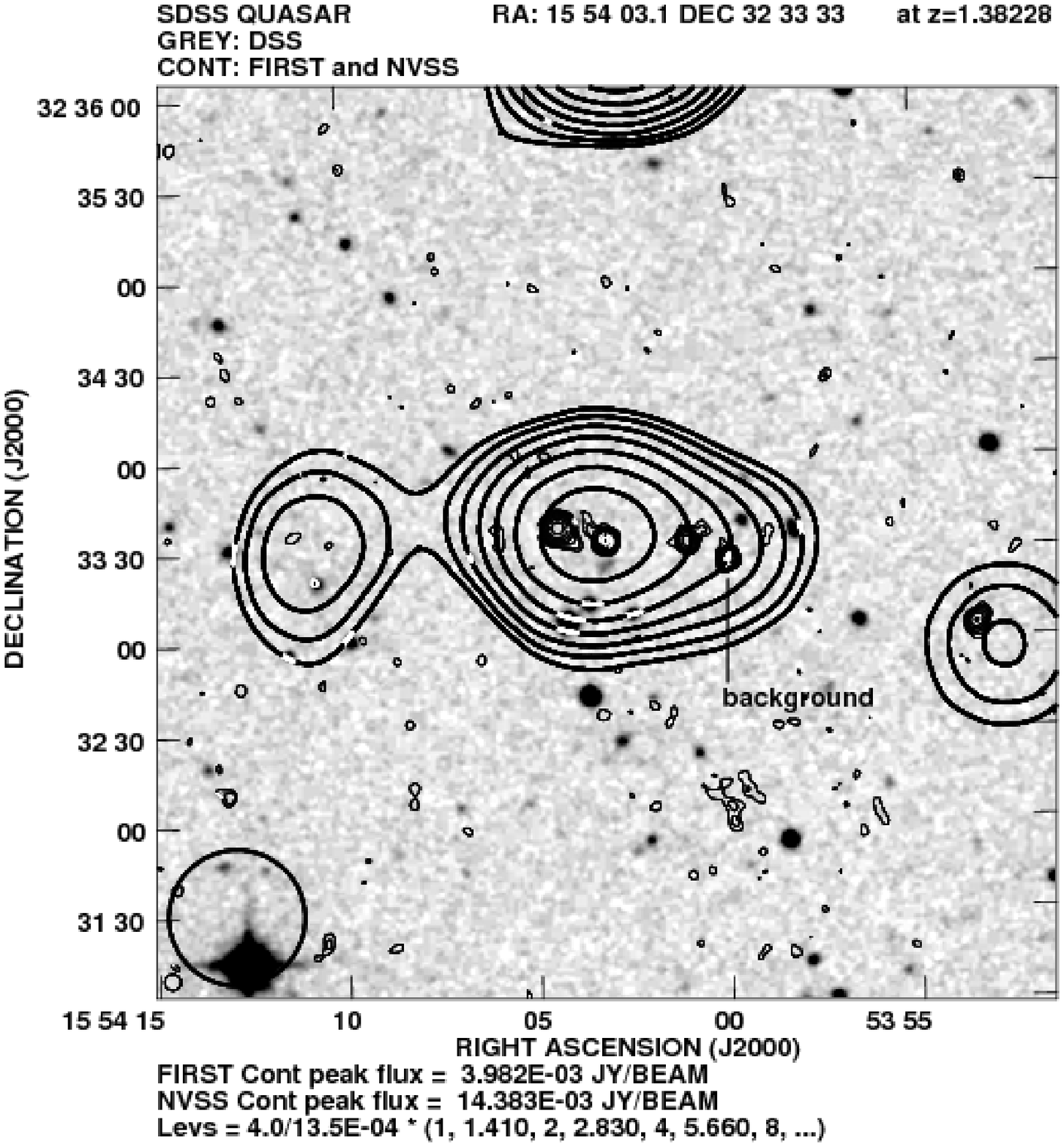,width=2.7in,angle=0}
            }
\caption[]{Candidate DDRQs J1215$-$0007 at a redshift of 0.99 (left panel) and J1554+3233 at a 
           redshift of 1.38228 (right panel) discussed in the text. A possible background source
           in J1554+3233 is marked.
          }
\end{figure}

\begin{figure}
      \hbox{
           \psfig{file=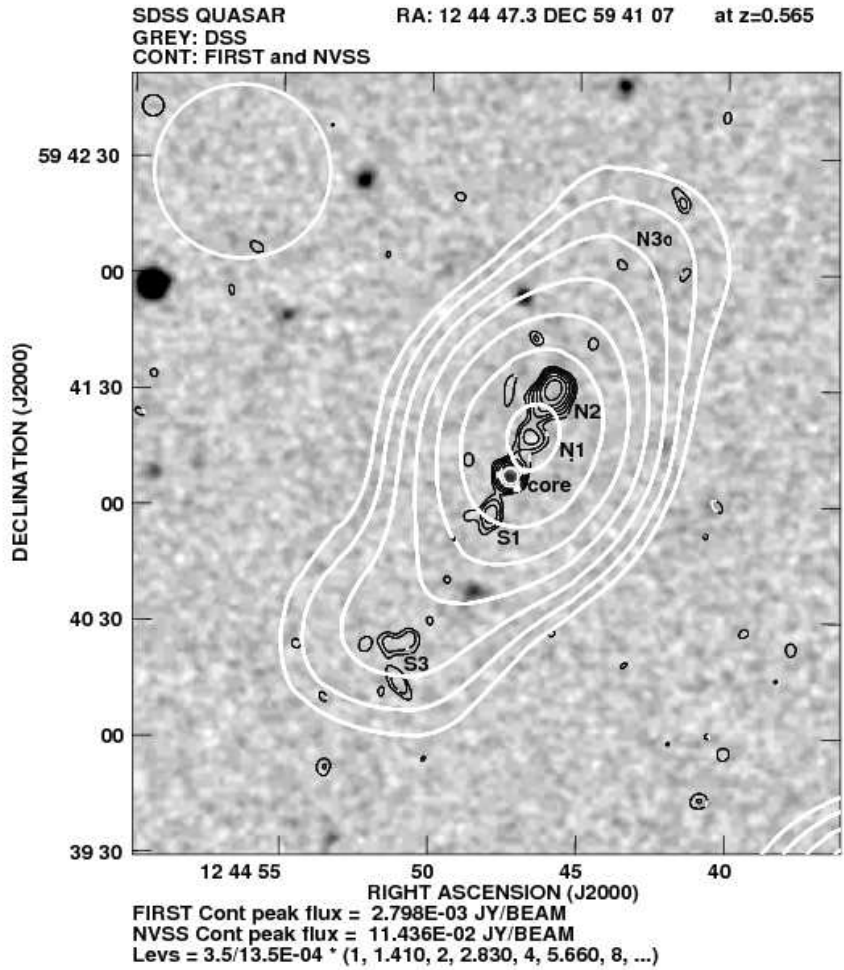,width=4.5in,angle=0}
           }
\caption{A quasar with possibly multiple cycles of AGN activity J1244+5941 at a redshift of 0.565
         discussed in the text. Multifrequency radio observations of different resolutions are required
         to explore this possibility.
         } 
\label{fig:J1244}
\end{figure}

\section{Candidate sources with recurrent activity}
In addition to the sources listed in Table 1, there are sources with features 
which have been suggested to be due to recurrent activity, but these are not
as clear as the sources listed in Table 1. To illustrate this we discuss a few cases.
As an example, consider the interesting source
B1144+352 (J1147+3501, B2 1144+35B) which has a Mpc-scale radio structure and where the arcsec-scale images show emission
on opposite sides of the nuclear or core component (Schoenmakers et al. 1999; Giovannini et al. 1999). The 
spectrum of the central component appeared peaked in Snellen et al. (1995), and was suggested to be a source with an
interrupted jet activity by Schoenmakers et al. (1999). However, the spectrum appears reasonably
flat when observed over a larger frequency range as seen in NASA Extragalactic Database (NED), and the source is also significantly
variable (Snellen et al. 1995; Schoenmakers et al. 1999; Giovannini et al. 1999), unlike the usual 
GPS sources (O'Dea 1998). In an optical study of bright flat-spectrum radio sources, Marcha et al. (1996)
suggest that it could be a diluted BL Lac object. VLBI observations of the central region show superluminal motion for the
approaching jet (Giovannini et al. 1999), and also motion in the counter-jet (Giovannini, Giroletti \&
Taylor 2007), suggesting an orientation of 33$^\circ$ of the jet axis to the line of sight. With the available
information the source appears to have a twin-jet structure in the central region rather than a young double,
as seen in J1247+6723 which has a GPS source in the central region. It would be interesting to image the source
with better sensitivity to trace the jet farther from the nuclear region, to try and clarify whether it might
be a source with recurrent activity.

Occasionally components are seen closer to the nucleus without a very well-defined edge-brightened structure.
Two examples of such cases are J0927+3510 and J1604+3438 which were suggested to be DDRG candidates by 
Machalski et al. (2006). However, later investigations showed these components closer to the core to have 
steep spectra, similar to that expected of the lobes. Their spectral ages were found to be consistent with 
the overall trend of age vs distance from the hotspot, except for the western component of J1604+3438 
(Konar et al. 2008; Jamrozy et al. 2008). Whether this component
might be due to recurrent activity or local re-acceleration of particles is unclear. 

So far only one possible DDRQ has been reported (Jamrozy, Saikia \& Konar 2009). 
In an effort to identify more DDRQs, we examined the radio structures, largely from the NRAO VLA Sky Survey 
(NVSS; Condon et al. 1998) and Faint Images of the Radio Sky at Twenty-Centimeters (FIRST; Becker, White \& Helfand 1995)
images, of quasars identified from the following surveys: Gregg et al. (1996), White et al. (2000),
Becker et al. (2001) and de Vries et al. (2006). Two of the possible DDRQ candidates (J1215$-$0007 and J1554+3233)
with their FIRST and NVSS contours superimposed on the DSS optical images are shown in Fig. 5,
while a similar image of a candidate triple-double radio quasar, TDRQ (J1244+5941) is shown in Fig. 6.  
J1215$-$0007 is associated with
a quasar at a redshift of 0.990 (1 arcsec=8.021 kpc) so that the size of the inner and outer doubles, 
are $\sim$90 and 1820 kpc respectively. The quasar J1554+3233 is at a redshift of 1.38228 (1 arcsec=8.493 kpc),
with one of the components being possibly associated with a faint optical object and marked as a background 
source in Fig. 5. If the outer components are related and thus possibly due to an earlier cycle of activity, the 
largest projected size would be close to 2 Mpc, while that of the inner double, defined by the two closest
companions, to the nucleus is $\sim$370 kpc. For the candidate TDRQ, J1244+5941, we show
the FIRST and NVSS contours superimposed on the DSS image of the field in Fig. 6. The compact 
component seen in the FIRST image is coincident with the position of the quasar which is at a redshift of 0.565
(1 arcsec=6.48 kpc). The NVSS image shows that the
outer lobes extend to about $\sim$970 kpc with the emission at the outer
edges being much weaker than in the central region. There is evidence of weak
emission at the outer edges in the FIRST image. These regions which represent
emission possibly from the earliest cycle of activity have been marked as N3 and
S3 in the image respectively. N2 has a prominent hotspot while the corresponding
feature towards the south is not visible in the FIRST image. The two closest components
to the core have been labelled as N1 and S1 and possibly represent emission from the most
recent cycle of activity. These features are unlikely to be knots in a jet since quasars
are known to have highly asymmetric or one-sided jets due to the effect of relativistic
motion. Clearly deep optical images of the field, along with multifrequency radio observations
with different resolutions are required to further clarify the nature of these sources.

\section{Considerations for models of double-double radio sources}
The existence of these sources with episodic activity clearly demonstrates that the 
AGN activity is not continuous over the life time of the radio source. It is not 
well understood whether the episodic activity may be related to the supply of fuel to 
the SMBH, collimation of outflows to form radio jets and/or instabilities in the jet 
production mechanism, and instabilities in the accretion disk. 

\noindent
{\bf Relic cocoons and propagation of jets:}
In most of the sources the emission from the different cycles of activity are reasonably
well aligned, so that the jets from the recent cycle would traverse through
the relic cocoon of the earlier cycle rather than the general intergalactic
or intragroup medium. For a few DDRGs, the inner double is of sub-galactic dimensions
and immersed in the interstellar medium of the host galaxy. 
For example, in 3C293, which also happens to be the DDRG showing the 
largest misalignment between the inner and outer doubles ($\sim$35$^\circ$), the size of the 
inner double is only about a couple of kpc, and the jets are traversing 
through the interstellar medium of the host galaxy as in other CSS and GPS objects.
The relic cocoon consists of magnetized relativistic
plasma, and is expected to be much lighter than the general interstellar or intergalactic
medium. Clarke \& Burns (1991) and Clarke (1997) have presented simulations of the
propagation of restarted jets in 2 and 3 dimensions respectively through such a medium. 
Clarke (1997) notes that the cocoon will end up being less dense and hotter than the jet 
material itself by a factor
of few, and the jet will propagate somewhat ballistically into the relic lobe. The
restarted radio jet is expected to be preceded by an observable bowshock, but this has not
been observed in any of the double-double radio sources. Clarke (1997) has discussed the
possibility of the Mach number being high enough ($>$1000) so that it hugs the jet along
its length making it difficult to distinguish from the jet, but suggested that such high
Mach numbers are unlikely. The inner doubles usually appear as edge-brightened lobes, with
a wide range in surface brightness of the hotspots, and
it is possible that these cocoons may have thermal densities higher than considered in
the simulations. Kaiser, Schoenmakers \& R\"ottgering  (2000) suggested the possibility of 
entrainment of dense
warm clouds from the intergalactic medium into the relic cocoon plasma, which could 
explain the observed structure of the inner lobes. However, although the time scales for      
entrainment and dispersion within the cocoon may be appropriate for the giant radio 
sources, the possibility of such a model being applicable for smaller sources needs to be
examined. It has also been speculated that the inner jets can reach the old cocoon 
boundary, cross it and then overtake the outer lobes, 
(see Kaiser, Schoenmakers \& R\"ottgering 2000; Marecki \& Szablewski 2009), 
with PKS 0349$-$27 (Morganti, Killeen \& Tadhunter 1993) being suggested to be a possible
example by Marecki \& Szablewski (2009). 
However, observationally, it may be difficult to unambiguously distinguish these from 
features due to backflow from the hotspots.

The properties of the relic cocoon in which the inner doubles are propagating outwards
may be probed more directly via X-ray observations, and also via radio polarization
measurements and, in principle, the symmetry parameters of the inner and outer doubles.
While X-ray emission has been detected as for example from the hotspots of 4C29.30
(Jamrozy et al. 2007), there has been no detection of X-ray emission from the relic
cocoon plasma in a DDRG. The X-ray emission in 4C29.30 has been suggested to be due to a combination of both 
non-thermal and thermal components, the latter being possibly due to gas shock heated
by jets from the host galaxy. The rotation measures which have been determined for only
a few of the DDRGs tend to be low. For example in the case of J1548$-$3216, the rotation
measure of the entire inner double is $-$11 rad m$^{-2}$ (Safouris et al. 2008), which
is close to the value of $-$14 rad m$^{-2}$ for the outer lobes (Saripalli et al. 2003).
The rotation measure of J0116$-$4722 also appears to be small with a mean of 3 rad m$^{-2}$
and a 1-$\sigma$ spread of 4 rad m$^{-2}$ (Saripalli et al. 2003). While these two are
about a Mpc in size or larger, polarization observartions of 4C29.30 show depolarization
near regions of optical line-emitting gas but the overall rotation measure is small
and believed to be largely of Galactic origin (van Breugel et al. 1986; Jamrozy et al. 2007). 
The rotation measures of the different components of J1835+6204 are within a few radians
of 58 rad m$^{-2}$ and are possibly of Galactic origin (Schoenmakers et al. 2000b). Although 
they generally tend to
be small, it would be useful to determine systematically the rotation measures of all the
double-double radio sources.

A comparison of the symmetry parameters of the outer and inner doubles, showed the
inner doubles to be more asymmetric in both brightness and location relative to the
radio core, compared with the outer doubles (Saikia, Konar \& Kulkarni 2006). This is unlikely to
be due to  higher velocities of the inner doubles, since the asymmetries are sometimes
not in the same sense as those of the outer doubles. While it needs to be confirmed
from a larger sample, it could reflect environmental asymmetries due to entrainment of 
material from the intergalactic medium (e.g. Kaiser, Schoenmakers \& R\"ottgering 2000), 
or intrinsic asymmetries in the oppositely-directed radio jets.

\begin{table}[h]
\caption{Multifrequency studies of DDRGs with GMRT and the VLA}
\begin{tabular}{ll   cc  cc c }
\hline
Source       & Alt.            & Spectral  & Spectral   & $\alpha_{\rm inj}$ & $\alpha_{\rm inj}$ & References \\
             & name            &age (outer)& age (inner)&  (outer)           &   (inner)          &          \\
             &                 &  Myr      &    Myr     &                   &                    &          \\
\\
    (1)      &    (2)          &  (3)     &  (4)      &  (5)       &    (6)             &     (7)  \\
\hline
J0840+2949   &  4C29.30        & $>$200     & $<33$   &  $\sim$0.8         & $\sim$0.8          &  1 \\
J1453+3308   &  4C33.33        & 47$-$58    &   2     &   0.57                &       0.57      &  2 \\
J1548$-$3216 &  PKS 1545$-$321   & 65$-$75    & $\sim$5$-$14  &   0.58, 0.54          & 0.61            &  3 \\
\hline
\hline
\end{tabular}
1: Jamrozy et al. (2007); 2: Konar et al. (2006); 3: Machalski, Jamrozy \& Konar (2010)  
\end{table}

\noindent
{\bf Time scales:} To understand the time scales between the cessation and renewal of AGN or 
jet activity, attempts have been made to estimate the spectral and dynamical or kinematic
ages of the outer and inner doubles in the DDRGs. One has to bear in mind the known caveats
while estimating spectral ages (see Blundell \& Rawlings 2001 for a comprehensive critique;
Jamrozy et al. 2008 for a brief discussion), and that
dynamical age estimates are usually larger than spectral ages by a factor of a few (see
Machalski, Jamrozy \& Saikia 2009), possibly also due to reacceleraton of particles in the lobes.
To illustrate the range of time scales which have been suggested let us consider a few cases.
Three of the DDRGs have been studied using both GMRT and VLA observations to determine their spectra
over a large frequency range (Table 2). The difference in spectral ages between the inner and outer
doubles range from $\sim$5$\times$10$^7$ yr to $\sim$10$^8$ yr. Using dynamical age estimates the
difference is likely to be larger, as seen
in the case of J1548$-$3216, where the dynamical ages of the outer and inner lobes have been
estimated to be $\sim$132 and 9 Myr respectively. For the sample of 5 DDRGs modelled by Kaiser,
Schoenmakers \& R\"ottgering (2000), the difference in ages between the outer and inner doubles is $\approx$10$^8$ yr.
In the case of J0116$-$4722, Saripalli, Subrahmanyan \& Udaya Shankar (2002) estimate that the
previous activity phase may have stopped at most 7$\times$10$^7$ yr ago, while the current activity
phase may have commenced 3$-$5$\times$10$^6$ yr ago assuming a hotspot speed of 0.2$-$0.3c. 
Safouris et al. (2008) estimate a time scale of 5$-$10$\times$10$^5$ yr between the cessation and
renewal of jet activity for J1548$-$3216 using their time scale limit for hotspot expansion. 
The presence of a hotspot in one of the lobes has been used to constrain the time scale of
interruption to less than a few Myr in the case of J0935+0204 (Jamrozy, Saikia \& Konar 2009)
and J1835+6204 (Schoenmakers et al. 2000b). In the case of Cygnus A, where there is evidence of
relic plasma, the time scale between successive episodes of jet activity has been constrained
to be $\sim$10$^6$ yr (Steenbrugge, Blundell \& Duffy 2008; Steenbrugge, Heywood \& Blundell 2010). 
There is a wide range of time scales inferred from the existing observations,
and it is possible that it might be difficult observationally to distinguish significantly smaller
time scales of episodic activity.

\noindent
{\bf FRI/FRII dichotomy:} It is interesting to note that some of the inner doubles in DDRGs
has an FRII structure although its luminosity belongs to the FRI category or lie in the 
borderline of the FRI/FRII classification. To explain the FR dichotomy (Fanaroff \& Riley 1974;
Baum, Zirbel \& O'Dea 1995; Zirbel 1997) there have been two broad classes of explanations: intrinsic
ones involving a fundamental difference in the SMBH or jet properties, and extrinsic
ones attributing it to interactions of the jets with the external media (see Bicknell 1984, 1995;
Wilson \& Colbert 1995; Reynolds et al. 1996; Meier 1999; Gopal-Krishna \& Wiita 2000).   
The edge-brightened structures of both the inner and outer doubles in these sources which have 
different luminosities and are propagating through different environments, suggest that the 
FRI/FRII difference might be due to differences in the central engine. 

\noindent
{\bf Injection spectral indices:}
Low-frequency observations of both the inner and outer lobes along with high-frequency ones
give the best possible estimates of the injection spectral indices, $\alpha_{\rm inj}$, which refer to 
the radiating
particles accelerated in the vicinity of the hotspots. So far, the three sources we have studied
using both GMRT and VLA observations to cover a reasonably large frequency range (Table 2), suggests
very similar injection spectral indices for the inner and outer doubles. It is interesting that this
is the case, although the external environments with which the jets are interacting leading to the 
development of shocks and acceleration of particles could be different. This needs to be examined
for a  larger sample of sources.

\noindent
{\bf Variability of cores:}
Normally, core-dominated sources are strongly variable, while weak cores in lobe-dominated sources
are at best expected to be mildly variable. This can be understood in terms of relativistic jets
which are inclined at small angles to the line of sight in the case of core-dominated sources
(e.g. Blandford \& K\"onigl 1979), while lobe-dominated sources are believed to be inclined at
large angles to the line of sight (e.g. Orr \& Browne 1982; Kapahi \& Saikia 1982; Barthel 1989).
The cores of two of the DDRGs, where multi-epoch observations have been put together, namely 4C29.30 
(Jamrozy et al. 2007) and J1453+3308 (Konar et al. 2006) show evidence of significant variability.
Again, in order to understand whether this might reflect a more general trend for double-double
radio sources requires observations of a larger sample.

\begin{figure}
      \hbox{
            \psfig{file=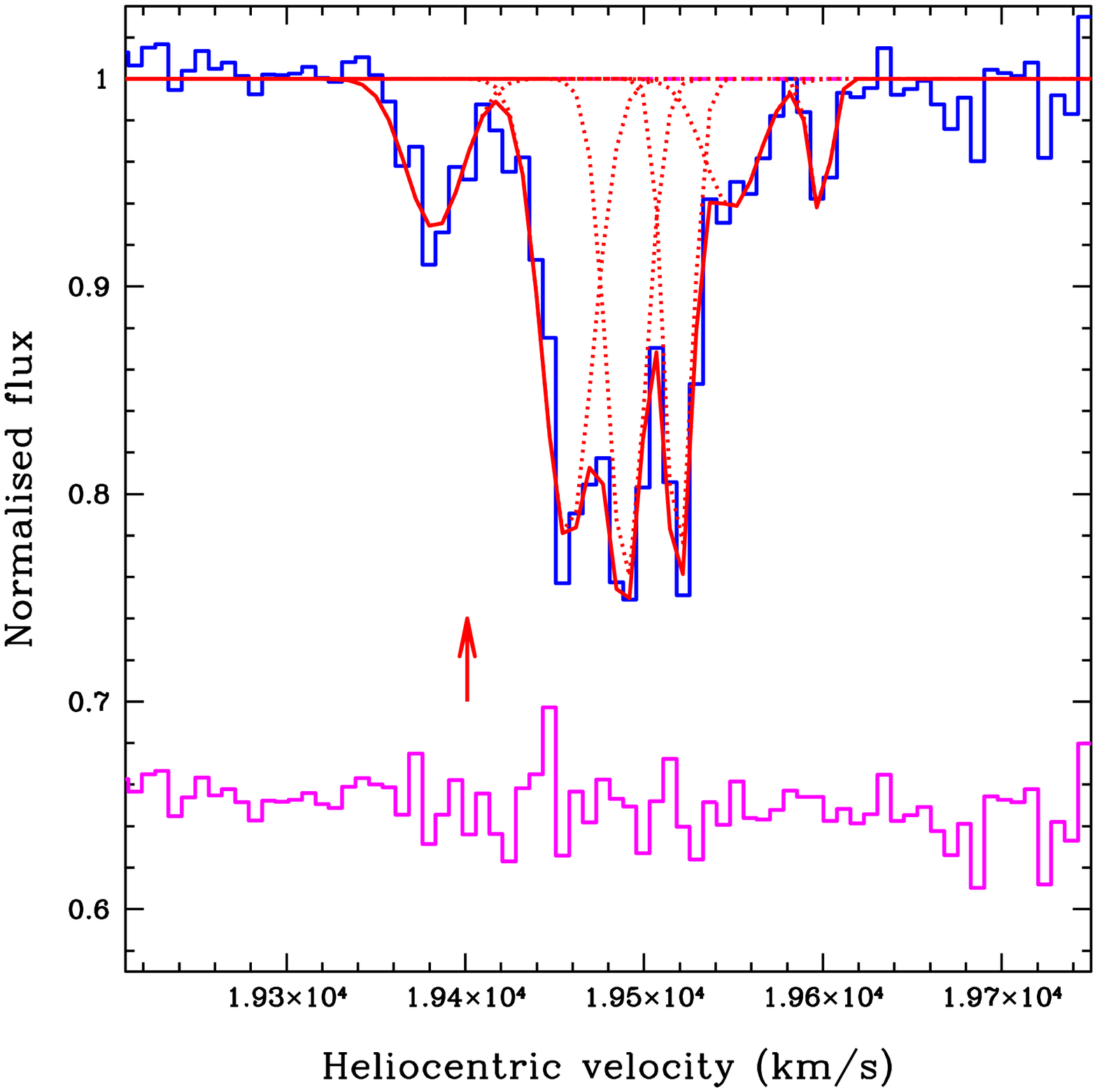,width=2.7in,angle=0}
            \psfig{file=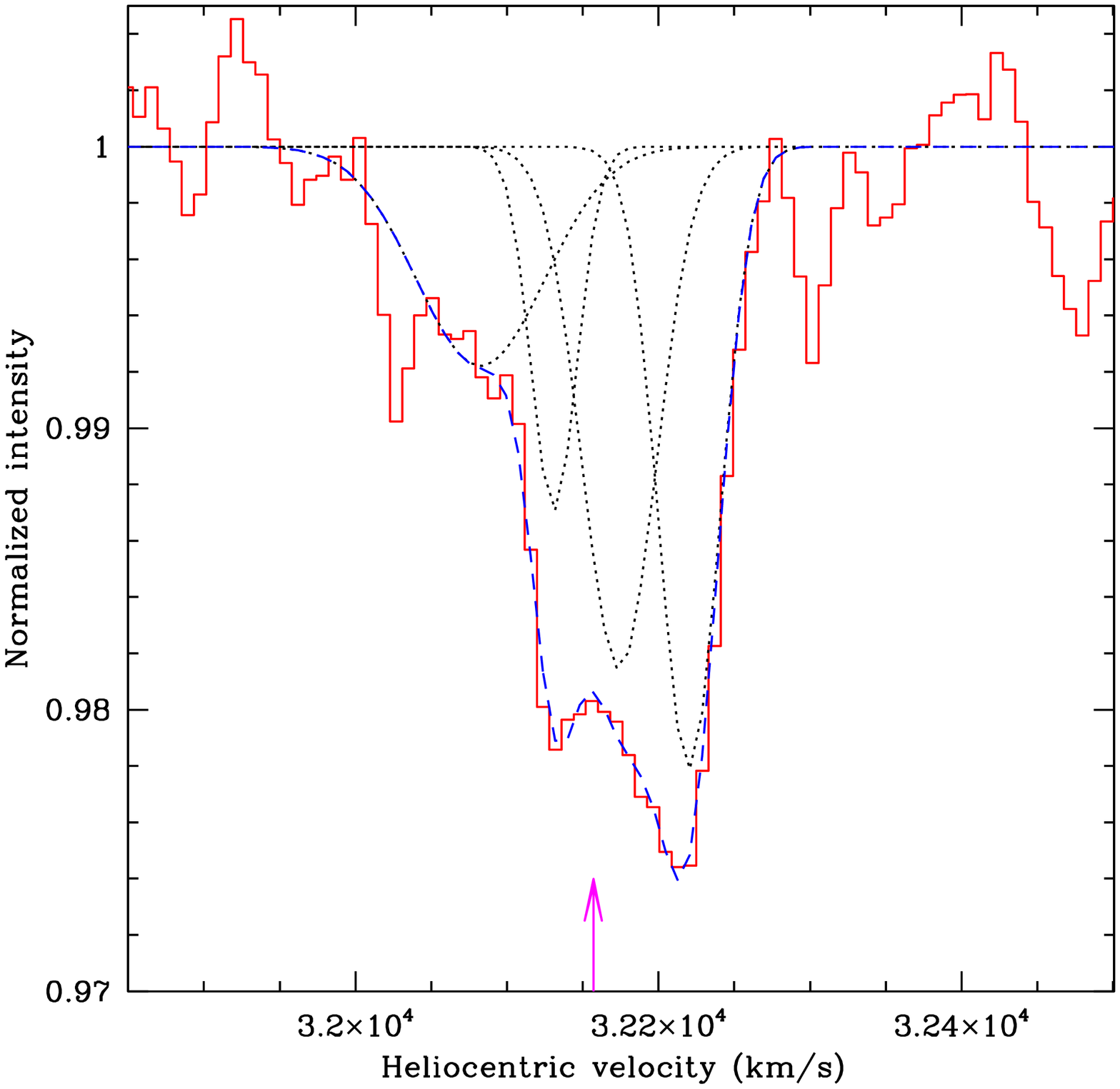,width=2.7in,angle=0}
            }
\caption[]{The H{\sc i} absorption spectra towards the core of 4C29.30 (left panel) and 
            the central GPS source in J1247+6723 (right panel). The Gaussian components and
            the sum of these compoents are also shown. In the left panel, the histogram at
            the bottom shows the residuals shifted upwards. The arrows mark the systemic
            velocity as measured from the optical emission lines. The spectra have been
            reproduced from Chandola, Saikia \& Gupta (2010) and Saikia, Gupta \& Konar (2007)
            respectively. 
          }
\end{figure}

\section{Rejuvenation of activity and supply of gas}
\label{sec:result}
If the rejuvenation of activity is due to a fresh supply of gas, one might be able find evidence
of this gas via either observations of CO (Saripalli \& Mack 2007) or H{\sc i} gas towards the
central regions of these objects. 
A millimeter-wave (CO) spectroscopic survey of an IRAS flux density-limited sample of 33 radio galaxies 
showed detections in 9 of these objects yielding a detection rate of $\sim$25 per cent. The molecular
gas masses of the detected galaxies are in the range of $\sim$(0.4$-$7)$\times$10$^7$ M$_\odot$, and 
eight of the nine molecular
gas-rich radio galaxies have close companions and/or exhibit tidal features (Evans et al. 2005). It
is also relevant to note that only one FRII galaxy out of 12 was detected, the remaining ones being either
FRIs or with compact radio jets. For a small sample of 9 radio galaxies with evidence of rejuvenated
radio activity which was observed for CO with the IRAM 30-m telescope (Saripalli \& Mack 2007), only 
one, namely 3C293, was known to have abundant molecular gas (Evans et al. 1999). This was also the FRII 
radio galaxy with a CO detection, suggesting that the detection rate for these rejuvenated radio sources 
are consistent with that for FRII radio galaxies. 
Ocana Flaquer et al. (2010) also found a similar lower detection rate for FRII galaxies, and also noted the gas 
mass to be higher for FRII galaxies, but suspected that this may be due to a Malmquist bias.

Saikia, Gupta \& Konar (2007) explored any possible relationship between rejuvenation of radio 
or jet activity and the occurrence of H{\sc i}. Unfortunately the number of sources
is still small because most of the rejuvenated radio sources have weak radio emission 
in the central or nuclear region. Therefore, the observations have so far been done 
on sources with the flux density within a few kpc of the nuclear region being larger
than $\sim$100 mJy.  The DDRG 3C236 which is a giant radio galaxy with a projected
linear size of $\sim$4250 kpc, and a compact steep-spectrum source in the nuclear region,
shows evidence of star formation and 
H{\sc i} absorption against a lobe of the inner radio source (Conway \& Schilizzi 2000; 
Schilizzi et al. 2001; O'Dea et al. 2001). H{\sc i} absorption has been reported towards
the central source of the giant radio galaxy J1247+6723, which is a GPS source and has
a compact double-lobed structure, by Saikia, Gupta \& Konar (2007).  
An interesting case is 3C293, which is a 
misaligned DDRG, and exhibits absorption features both blue- and red-shifted relative
to the the systemic velocity
(Beswick, Pedlar \& Holloway 2002; Beswick et al. 2004), and fast outflowing gas
blue-shifted by upto $\sim$1000 km s$^{-1}$ (Morganti et al. 2003; Emonts et al. 2005).
The absorption profile towards the central component of the rejuvenated radio galaxy
4C29.30 (Jamrozy et al. 2007),
is well resolved and consists of six components; all but one of which appear to be
red-shifted relative to the optical systemic velocity (Chandola, Saikia \& Gupta 2010).
Another case is Cen A, where blue-shifted H{\sc i} absorption is seen towards the central 
region in addition to the broad red-shifted gas which may be a blend of several components
within a 100 pc region (e.g. Sarma, Troland \& Rupen 2002; Morganti et al. 2008). The 
H{\sc i} absorption has been ascribed to a cold, circumnuclear disk (Morganti et al. 2008).

While the number of sources is still extremely small, the detection of absorbing H{\sc i}
gas in the rejuvenated galaxies appears to be more frequent than for CSS and GPS objects
(Vermeulen et al. 2003; Gupta et al. 2006). The detection rate for the GPS objects 
listed by Gupta et al. (2006), which has a higher H{\sc i} detection rate than for CSS and
larger objects, is $\sim$45 per cent. 
Considering the H{\sc i} column densities listed by Gupta et al. (2006),
the median value for GPS sources is $\sim$3$\times$10$^{20}$ cm$^{-2}$. The rejuvenated
radio galaxies listed earlier have column densities in the range of $\sim$8$-$50$\times$10$^{20}$ 
cm$^{-2}$. The line profiles are also generally complex with mulitple components. Clearly 
an effort is required to make observations of the sources with weaker cores and other
rejuvenated sources to re-examine this trend.

\section{Fossil record of episodic AGN heating in clusters}
One of the promising possibilities proposed to solve the `cooling flow problem' in clusters
of galaxies, i.e. the lack of evidence of gas cooling to low temperatures as predicted by
the models (Peterson \& Fabian 2006) is feedback by the AGN. The importance of AGN feedback
was highlighted as early as 1990 by Pedlar et al. (1990) who estimated that the energy input
is comparable to the X-ray luminosity over the central region of the cluster, and they noted that ``its effects 
should be considered in models of cooling inflows in the central parts of the Perseus cluster.''
A remarkable manifestation of this is for the jets from the AGN to inflate lobes of radio-emitting
plasma, which pushes aside the X-ray emitting gas in the cluster leaving apparent cavities in
the X-ray images (McNamara \& Nulsen 2007; Gitti et al. 2009 and references therein; B\^irzan et al.
2009 and references therein). Besides addressing the cooling problem, the energy that the cavities
release in the intracluster medium could also affect the process of galaxy formation and evolution
(e.g. Croton et al. 2006). Images from the {\it Chandra} observatory have shown that many clusters
have X-ray cavities in their atmospheres, and a combination of deep X-ray and low-frequency radio
imaging are likely to provide the most promising diagnostics for determining the ages, energetics
and duty cycles of AGN in these clusters. The mechanical energy required to inflate these bubbles
is $\sim$10$^{61}$ ergs, and the rise times of the bubbles due to buoyancy is $\sim$50$-$300 Myr,
indicating that these bubbles contain the fossil record of AGN activity in the cluster.

\begin{figure}
      \hbox{
            \psfig{file=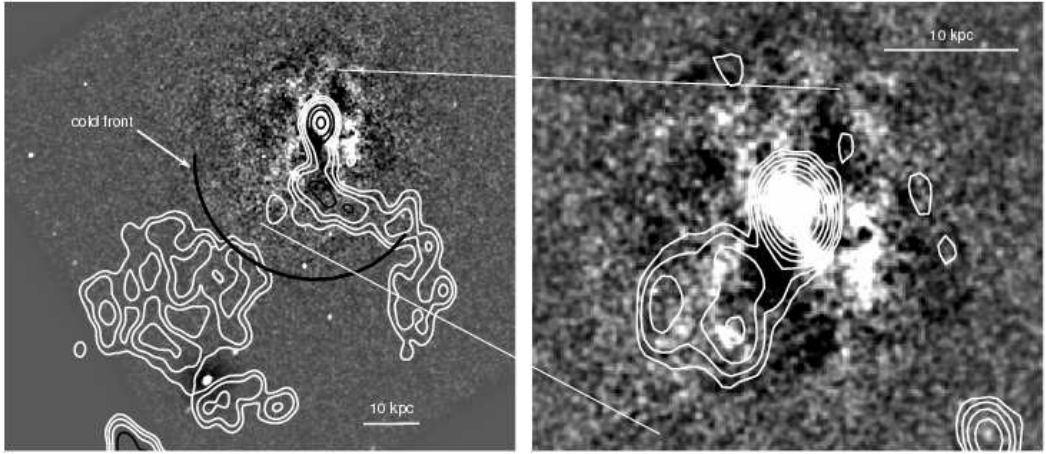,width=5.4in,angle=0}
            }
\caption[]{GMRT 235-MHz (left panel) and 610-MHz (right panel) contours overlaid on the 
           {\it Chandra} unsharp masked image of the X-ray bright group NGC5044 in the 0.3$-$2 keV band 
            (David et al. 2009). 
           This figure has been reproduced from Giacintucci et al. (2009).
          }
\end{figure}

As an illustration of the importance of combining deep X-ray and low-frequency radio imaging we
show the results of NGC5044 presented by Giacintucci et al. (2009), which suggest two or possibly
three cycles of AGN activity. The {\it Chandra} observations reveal that the group hosts many
small radio quiet cavities, filaments and a semi-circular cold front (David et al. 2009). The GMRT
610-MHz image (Fig. 8, right panel) shows a core and a lobe which extends along a cold filament,
while the 235-MHz image (Fig. 8, left panel) shows not only more extended emission in the central
region, but a detached radio lobe. The western edge of this detached lobe is close to the cold front,
and has possibly been produced in an earlier cycle of activity. This feature is not seen in the 
610-MHz image, suggesting that it has a steep spectrum as expected for an old lobe (Giacintucci et al. 2009). 
The GMRT observations reveal at least two cycles of AGN activity, the youngest is seen close to the core
and visible in the 610-MHz image and the oldest is identified with the detached radio lobe. Detailed
spectral and kinematical studies of such features would be invaluable to estimate time scales of episodic
activity in AGN.
  
\section{Concluding remarks}
Deep X-ray and low-frequency radio studies have proved to be useful to probe episodic or recurrent
AGN activity in radio galaxies, as seen in the case of Cygnus A (Steenbrugge, Blundell \& Duffy 2008;
Steenbrugge, Heywood \& Blundell 2010), as well in clusters of galaxies as seen in the case of the 
X-ray bright group NGC5044 (e.g.  David et al. 2009; Giacintucci et al. 2009).
The existing observations have, however, possibly revealed only the tip of the proverbial iceberg. The
GMRT observations at low frequencies have been extremely valuable in identifying emission from earlier
cycles of activity (e.g. David et al. 2009; Giacintucci et al. 2009) as well as determing the low-frequency
spectra of both the inner and outer lobes of emission in double-double radio sources (e.g. Konar et al. 
2006; Jamrozy et al. 2007; Machalski, Jamrozy \& Konar 2010). However, the surface brightness sensitivities 
achieved in many of these studies could be better. For example, in a study of the ELAIS-N1 field at 325 MHz 
using the GMRT, the rms noise in the image was $\sim$50 $\mu$Jy beam$^{-1}$ (Sirothia et al. 2009b). 
Deeper low-frequency images should enable us to identify many more sources with clear signatures of
episodic activity, although attempts by Sirothia et al. (2009a) for largely small sources did not yield
any new unambiguous candidates. In this context it is also relevant to identify relic lobes in 
radio sources and examine evidence of more recent activity. One of the earliest examples of a `dying' radio 
source, B2 0924+30, associated with the galaxy IC 2476, was reported by Cordey (1987), while a sample of similar 
radio galaxies with steep spectra but no current AGN activity, compiled from VLA and GMRT observations, has 
been reported by Dwarakanath \& Kale (2009). Godambe et al. (2009) have presented a study of three giant
radio sources (J0139+3957, J0200+4049 and J0807+7400) with possible relic lobes, but which have detected
radio cores. They have discussed the possibility that the lobes are due to an earlier cycle, while the cores
represent more recent activity, but this needs to be examined further.
With the advent of new low-frequency telescopes such as the
Low Frequency Array (LOFAR), Long Wavelength Array (LWA), Murchison Widefield Array (MWA), Australian Square Kilometre 
Array Pathfinder (ASKAP), the Karoo Array Telescope (known as MeerKAT) and on a somewhat longer time scale, the
Square Kilometre Array (SKA),  we are in an exciting phase of studying episodic activity in AGN.

It is relevant to note that 7 of the 22 sources listed in Table 1 have a redshift $<$0.1 and 14 have
a redshift $<$0.2. The highest spectroscopic redshift in the sample is 1.779 for 3C294 for which recurrent 
activity has been inferred from radio and X-ray emission possibly due to ICCMB photons. This is followed by 
0.5194 for a galaxy (J1835+6204) and 0.6491 for 
a quasar (J0935+0204). Essentially most sources with episodic activity are at moderate or at low
redshifts. Although the number of sources with signs of recurrent activity (Table 1) span a range of
linear sizes from approximately a 100 kpc to over a Mpc, many of the classic double-double radio sources
are found in giant radio sources, which are defined to be over about a  Mpc in size (e.g. Ishwara-Chandra
\& Saikia 1999; Schoenmakers et al. 2001). It is interesting to note that there is a paucity of giant radio 
sources at large redshifts. Although the highest redshift giant source associated with a galaxy is at a redshift of 1.88
(Law-Green et al. 1995) and a few associated with quasars between redshifts of $\sim$1 and 1.7 
(Singal, Konar \& Saikia 2004; Kuligowska et al. 2009),
most giant radio sources are at small or intermediate redshifts, less than $\sim$0.5 (e.g. Ishwara-Chandra \& Saikia 1999; 
Schoenmakers et al. 2001). The deficit of giant sources at large redshifts may be due to radiative
losses due to inverse-Compton scattering with the cosmic microwave background radiation, which does appear
to affect the strength of the bridge emission (e.g. Gopal-Krishna, Wiita \& Saripalli 1989;
Konar et al. 2004). ICCMB X-rays have been directly observed and shown to be an excellent tracer of aged 
synchrotron plasma in the giant radio source 6C 0905+3955 (Blundell et al. 2006). 
While examining the evolution of radio sources, and the deficit of giants at large redshifts,
issues related to youth-redshift degeneracy have to be also borne in mind
(see Blundell, Rawlings \& Willott 1999; Blundell \& Rawlings 2000). In the coming years, deep low-frequency imaging with
the GMRT, LOFAR and the other upcoming telescopes may help identify giant sources over a 
larger range of redshifts and scales,
as well as help identify relic lobes at Mpc scales due to earlier cycles of activity. Deep X-ray imaging
may also help reveal these relic emission, as seen in the cases of Cygnus A (Steenbrugge, Blundell \& Duffy 2008),
and 3C294 (Erlund et al. 2006), and possibly in 3C191 where the X-ray emission extends well beyond the radio
contours (Erlund et al. 2006), and help open up another major window to study recurrent activity in AGN.

Although the known double-double radio sources tend to have low rotation measures, which have been 
usually inferred from high-frequency ($>$1 GHz) measurements, low-frequency polarization observations
may help us to probe the properties of the cocoon. Deep X-ray observations of the double-double radio
sources would also be invaluable to probe the properties of the ambient medium, and understand the 
formation of shocks and hotspots in the inner doubles. Luminous inner doubles with prominent hotspots in sources such as 
J1006+3454 (3C236), J1247+6723 and J1352+3126 (3C293), which are of subgalactic dimensions, can be understood  
in terms of interaction of the jets with the interstellar medium of the host galaxy. However, a source such
as J0041+3224 which has prominent inner hotspots with the inner double being more luminous than the outer
one at 1400 MHz, and a size of $\sim$170 kpc for an estimated redshift of 0.45, could challenge our 
conventional understanding. The inner double appears to extend well beyond the confines of the host galaxy, with the
caveat that its
redshift needs to be confirmed spectroscopically. The similarity of the injection spectral indices for the
inner and outer doubles, which possibly propagate through significantly different media also needs to 
be explored from further observations. If confirmed, it raises the possibility that injection spectral indices
may be set close to the central engine.

Another useful approach, which we have not touched upon in this review, would be to explore the star formation history
in sources with signs of episodic AGN activity in an effort to understand possible relationships between
these two forms of activity in the circumnuclear regions of the host galaxies. Clarificaton of these
aspects may have significant consequences in our understanding of AGN feedback, growth of black holes, and 
galaxy formation and evolution (e.g. Best et al. 2006; De Young 2010).  

\section*{Acknowledgments}
Different pieces of work covered in this review have been done in collaboration with Yogesh Chandola, 
Sagar Godambe, Neeraj Gupta, C.H. Ishwara-Chandra, Nimisha Kantharia, Chiranjib Konar, Vasant Kulkarni, 
Jerzy Machalski, Karl-Heinz Mack, Aneta Siemiginowska, Sandeep Sirothia,
{\L}ukasz Stawarz and Paul Wiita. We thank them for their contributions over the years.
We thank Simona Giacintucci for permission to use Fig. 8, and Katherine Blundell, Gopal-Krishna, Neeraj Gupta,
Ananda Hota, Andrzej Marecki, Jan Vrtilek and Paul Wiita for their detailed and very helpful
comments on the  manuscript, which were given at very short notice. DJS thanks ICRAR, UWA for a Visiting 
appointment while this review was being written. 
MJ acknowledges the MNiSW funds for scientific research during the years 2009$-$2012 under 
contract No 3812/B/H03/2009/36.
This research has made use of the NASA/IPAC extragalactic database (NED)
which is operated by the Jet Propulsion Laboratory, Caltech, under contract
with the National Aeronautics and Space Administration.  This research has made use of NASA's Astrophysics Data System.

\end{document}